\documentclass[twocol]{ametsocV6.1}
\usepackage{amsfonts,amsmath}

\title{{The impact of two-dimensional filtering on SWOT observations with white noise}}
\authors{Ryan Sh\`iji\'e D\`u\aff{a},\correspondingauthor{Ryan Sh\`iji\'e D\`u, ryan.du@mines.edu}
Momme Hell\aff{b},
Luc Lenain\aff{c},
Fabrice Ardhuin\aff{d}, 
A. B. Villas Bôas\aff{a}
}

\affiliation{\aff{a}{School of Earth, Space and Planets, Colorado School of Mines, Golden, CO, USA}\\
\aff{b}{Woods Hole Oceanographic Institution, Woods Hole, MA, USA}\\
\aff{c}{Scripps Institution of Oceanography, University of California San Diego, La Jolla, CA, USA}\\
\aff{d}{Univ. Brest, CNRS, IRD, Ifremer, Laboratoire d'Oc{\'e}anographie Physique et Spatiale (LOPS), IUEM, Brest, France}
}
\abstract{
The Surface Water and Ocean Topography (SWOT) mission provides two-dimensional observations of sea surface height (SSH) at unprecedented spatial resolution, enabling exploration of ocean variability down to scales of $O(10~\mathrm{ km})$. At these scales, however, interpreting SSH variability is challenging because ocean dynamical signals overlap with measurement noise, and their respective spectral signatures are not yet fully understood.
Recent analyses of SWOT 2-km posting observations have shown that along-track spectra {transition to flatter but still red spectra at scales around 30 km. These flatter portions of the spectra have} a power-law-like behavior and spectral slopes {of approximately $-1$ or steeper, and} their magnitudes and slopes are correlated with SWOT measurement noise {magnitude}. 
Here, we investigate the hypothesis that {these flatter} but still red along-track {small-scale} spectra can arise from two-dimensional filtering and aliasing of spatially uncorrelated (white) noise. Using synthetic experiments, we show that the resulting one-dimensional along-track spectra exhibit {a similar transition to} red, power-law-like behavior at {scales of 15--50 km, qualitatively similar to spectral behavior reported in SWOT} observations.
The {transition scale and} apparent spectral slopes depend on the noise level, its cross-track variability, and the background ocean signal. 
This finding highlights the importance of carefully accounting for measurement noise and processing effects when interpreting SWOT spectra, and suggests that {a white} noise model should serve as a baseline null hypothesis for small-scale spectral analyses.
}

\begin{document}
\maketitle

\section{Introduction}
The Surface Water and Ocean Topography (SWOT) mission represents a major advancement in our ability to observe the ocean. It provides two-dimensional observations of sea surface height (SSH) at unprecedented spatial resolution, improving on traditional nadir altimeter observations \citep{ArcherEtAl_25}. The performance of SWOT exceeds the engineering requirements, allowing us to observe smaller scales than what was expected prior to launch \citep{PeralEtAl_24,Chelton_24}. 

High-quality global measurements from SWOT present exciting opportunities for oceanography, but, as with many new observing systems, they also extend into dynamic ranges that are not yet fully understood. Specifically, SWOT resolves SSH {signals associated with ocean physics} at small scales that are difficult to interpret, because at {the submesoscale $(O(10-100\text{ km}))$} the ocean consists of a complex \emph{submesoscale soup} of balanced motions {(eddies and fronts) and (unbalanced)} internal gravity waves \citep[IGWs,][]{McWilliams_16}. 
These turbulent ocean dynamics typically exhibit different spectral behaviors in the SSH field, such that for balanced motions we expect steep $k^{-4}$ or $k^{-5}$ power-law behavior \citep{Charney_71,HeldEtAl_95} {and for the IGW continuum} we expect shallower $k^{-2}$ power-law behavior  \citep{GarrettMunk_72,SamelsonFarrar_24}. 
SWOT can even phase-resolve long ocean surface waves of $O(1\mathrm{ km})$ \mbox{\citep{ArdhuinEtAl_25, VillasBoasEtAl_25a}}, which appear as local maxima in the spectra of SSH \mbox{\citep{VillasBoasEtAl_22,PeralEtAl_24}}.

To interpret SWOT's spectral slopes correctly, it is important to quantify the measurement noise, in particular at high wavenumbers, i.e., submesoscales, where noise might dominate the ocean dynamical signal \citep{WangEtAl_19,Chelton_24,PeralEtAl_24}. Recent studies indicate that measurement noise makes a major contribution to SWOT small-scale $(<50\mathrm{km})$ along-track spectra. For example, \citet{ZhangCallies_25a} analyzed the spectral behavior of global SWOT SSH observations collected during the CAL/VAL phase and found that, in regions where balanced signals dominate, the SWOT 2-km along-track spectra {transition to flatter but ostensibly red power-law spectra near 30 km}, with slopes steeper than $-1$. They also show that the magnitude and slope of these small-scale spectra are correlated with Significant Wave Height (SWH), which is {in turn correlated with} the magnitude of the SWOT SSH {measurement} noise \citep{PeralEtAl_24}. {Adding further complexity to the characterization} of SWOT measurement noise, \citet{KacheleinEtAl_25} showed that {its noise is expected to be higher at the swath edges, leading to changing} small-scale slopes across the swath, with flatter spectra at the swath edge. 
However, data and methodological uncertainties have thus far prevented us from determining whether the red spectra arise primarily from measurement noise or from ocean dynamics. 

The observed red behavior in SWOT small-scale spectra $(O(<50\mathrm{km}))$ is surprising and contributes to the difficulty of separating noise from signal.
{The power-law-like behavior of SWOT 2-km posting (i.e., data sampled on a grid with 2-km spacing) along-track small-scale $(<50\mathrm{km})$ spectra is visually striking, and it is tempting to treat these as true power-laws to diagnose and interpret their spectral slopes (cf. \citet{ZhangCallies_25a,KacheleinEtAl_25, SkinnerEtAl_25}).}
{However, we generally expect} SSH measurement noise to be white at small scales. In fact, \citet{PeralEtAl_24} show that, in the SWOT 250-m posting data, noise in the along-track spectra is white for scales larger than 2~km, which are not affected by the on-board processing \citep[OBP,][]{PeralOBP_21}. 
{To reconcile the noise behavior in SWOT spectra at the two data postings, we show how white noise in the 250-m posting data can appear as a red power law in the SWOT 2-km along-track spectra.}

{Specifically, we explore the potential behavior} of white noise in the along-track spectra from SWOT's 2-km posting product using a series of synthetic experiments. We start from 250-m posting data composed of a prescribed steep ocean dynamical signal plus two-dimensional white noise. 
The 2-km posting data are generated from this field by two-dimensional filtering {and posting}, and we investigate the resulting one-dimensional along-track spectra. This standard processing causes the small-scale along-track spectra to appear power-law-like and red. With our synthetic experiments, we show how the apparent slope depends on the noise level, cross-track noise variation, and background ocean dynamics.

The paper is organized as follows. Section \ref{sec:experiments} describes experiments {and analysis that produce ostensibly red power-law-like spectra which visually resemble $k^{-1}$ or steeper spectral decay} from simple ocean signals and white noise. We start with balanced ocean signals {in Section \ref{sec:experiments}\ref{sec:hamm_red}. 
Section \ref{sec:experiments}\ref{sec:theory} provides an analytical derivation of the results.}
Cross-track variation of SWOT noise is explored in Section \ref{sec:experiments}\ref{sec:cross}. The Garrett--Munk continuum is used for the ocean signals in Section \ref{sec:experiments}\ref{sec:GM}. Section \ref{sec:conc} concludes and discusses future work.

\section{{Experiment with SWOT-like data and white noise}}\label{sec:experiments}
{We use synthetic data that represent an idealized version of the SWOT sea surface height anomaly (SSHA) data. This includes contributions representing an ocean dynamical signal and measurement noise. All spectra plotted in this paper are one-sided physical power spectral densities in units of $\text{ cm}^2\text{cpkm}^{-1}$ estimated using ensemble and cross-track averaging.}

{Here, we restrict the ocean physics signals considered to balanced ocean dynamical features and the IGW continuum, which have spatial spectra with power-law behavior. We do not explore other ocean physics signals that could be a part of SWOT's data, such as coherent internal tides \citep{RayZaron_16,ZaronEtAl_26} or ocean surface wave swell \citep{VillasBoasEtAl_22, ArdhuinEtAl_25, VillasBoasEtAl_25a}, as they are not generally well represented by a single power law. We also exclude contributions to the mean sea surface, such as those associated with ocean topography \citep{YuEtAl_24a}, as we are focused on dynamical signals in the SSHA field.}

{We model SWOT measurement noise as uncorrelated (white) noise whose magnitude is determined by mean SWH and cross-track position. The noise is statistically independent of the simulated ocean dynamical signal. This is a simplification, as the SWOT measurement noise also depends on wind speed \citep{HuangWang_26}. For demonstration purposes, an SWH-dependent noise model, as presented in \citet{PeralEtAl_24} and explored by \citet{ZhangCallies_25a}, is sufficient.
We do not attempt to model the detailed physical mechanisms responsible for SWOT SSH measurement noise, as these remain an active area of research and are outside the scope of this paper \citep{ArdhuinEtAl_26,OsinaEtAl_26}. }

{All synthetic data are generated as Gaussian random fields. For the white noise component, this simply amounts to drawing independent Gaussian random variables at each 250-m posting location. For the ocean dynamical ``signal'', a Gaussian random field will miss many realistic features of ocean dynamics but does capture all second-order statistics of a stationary random field \citep{Yaglom_87a}. Since our analysis is only concerned with spectra in this manuscript, this representation is sufficient. We also note that Gaussian random fields are assumed for Gaussian process regression, which has been used for both nadir altimeter missions and SWOT applications, including denoising and interpolation \citep{Wunsch_96,UbelmannEtAl_21,ZhangCallies_25a, SkinnerEtAl_25}.}

{
Throughout this manuscript, we generate a large ensemble of independent samples, making the sampling uncertainty of the plotted ensemble mean spectra negligible. Spectral slopes in the synthetic experiments are estimated visually by comparison with the power-law reference lines shown in the figures. We do not fit power laws because the proposed filtered white-noise mechanism does not imply an underlying red-power-law process. Fitting a power law to these along-track spectra would therefore impose an unjustified assumed form on spectra. For our purposes, visual estimation is sufficient to demonstrate the qualitative behavior of the spectra.
}

\subsection{{Apparent red power-law-like spectra from filtering SWOT data with white noise}}\label{sec:hamm_red}
{In this section, we show that the observed $-1$ or steeper small-scale power-law-like slopes in SWOT data can arise from filtering and aliasing of two-dimensional white noise superposed on steep ocean spectra.}

\begin{figure*}
    \centering
    \includegraphics{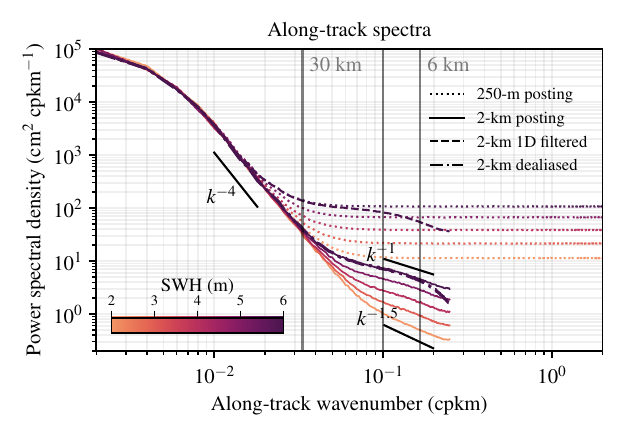}
    \caption{Along-track spectra of the synthetic SSH field for several noise levels corresponding to different significant wave heights (SWH, colors). {Spectra are averaged over cross-track position and a 200-member ensemble.} Dotted lines are along-track spectra for the 250-m posting data before Hamming filtering. Solid lines are along-track spectra for the 2-km posting data obtained by convolving {the 250-m posting field} with the 17-point two-dimensional Hamming filter used by SWOT {and sampling the result on the 2-km grid}. The darkest dash-dotted line is the dealiased two-dimensional Hamming-filtered version.
    The darkest dashed line shows one-dimensional along-track filtering applied to the same synthetic 250-m posting field ($\text{SWH}=6\text{ m}$ case). {The black solid power-law lines are for visual reference over the range where they are plotted.}}
    \label{fig:Hamm_red}
\end{figure*}

We generate {the synthetic} Gaussian random fields of size $50\text{ km} \times 500\text{ km}$ at 250-m posting, mirroring one swath of the SWOT ground track, which has a cross-track width of $50\text{ km}$. The spectrum of {the ocean signal component is} a two-dimensional Matérn kernel with a roughness parameter $\nu = 1.5$ \cite[p. 49]{Stein_99a}. Matérn kernels are used in turbulence research to represent power-law behavior at high wavenumbers, and the low-wavenumber transition prevents a singularity at the zero wavenumber {\citep{LillyEtAl_17a}. For spatial interpolation (e.g., optimal interpolation), the Matérn family provides a flexible covariance class that captures the local roughness of a random field \citep{Stein_99a}.}
{The full two-dimensional spectral density asymptotes to a slope of $-5$, while the azimuthally integrated isotropic spectrum and the along-track spectrum both asymptote to $-4$. This asymptotic behavior occurs for scales smaller than 100 km (see Fig. \ref{fig:Hamm_red})}, imitating the SSH fields of surface-buoyancy-driven balanced ocean motions \citep{HeldEtAl_95,ZhangCallies_25a,DuEtAl_25}.

At small scales, the field is dominated by noise uncorrelated in space, which has a white (zero-slope) along-track spectrum. 
We explore many representative noise levels, where higher SWH correlates with higher noise levels \citep{PeralEtAl_24}. 
Dotted lines in Fig.~\ref{fig:Hamm_red} show the along-track spectra {of the 250-m posting data} estimated by averaging over all tracks across the generated swath for 200 independent ensemble realizations. They follow the Matérn {ocean signal at low wavenumbers, while noise differentiates them at high wavenumbers.} For 2~m SWH, the {250-m posting along-track spectral noise level is based on re-scaling the SWOT prelaunch noise requirement, as explained below:}
\begin{align}
    2\times {\frac{2}{1/15}}\times\frac{1}{2.3^2} \text{ cm}^2\text{cpkm}^{-1}.\label{eq:noise_floor}
\end{align}
Here, $2 \text{ cm}^2\text{cpkm}^{-1}$ is the magnitude of the white-noise portion of the SWOT prelaunch noise requirement, which is specified for $\text{SWH}= 2\text{ m}$ and wavenumbers less than $1/15$ cpkm {\citep{SWOT_sci_req}}. {For spatially white two-dimensional noise, the one-dimensional along-track noise level is proportional to the retained cross-track wavenumber bandwidth. The factor {$2/(1/15)$ therefore scales the requirement, specified for cross-track wavenumbers up to $1/15$ cpkm, to the 250-m-posting Nyquist wavenumber of $2$ cpkm} (see Appendix A, \citet{WangEtAl_19}, and \citet{CheltonEtAl_22}). }
Finally, we account for the fact that the SWOT noise {root-mean-square amplitude is \(1/2.3\) times} the prelaunch requirement, according to preliminary studies on the SWOT noise \citep{Chelton_24,PeralEtAl_24}.
The noise for the higher SWH cases is scaled from the 2~m SWH case using the corresponding (2~m to 6~m SWH) total SWOT noise magnitude from the legend of Fig. 30a in \citet{PeralEtAl_24}. 
We acknowledge that {our experiments do not account for the OBP filtering} effect. 
{The OBP filtering} in the 250-m posting data does not significantly change our results, since the additional Hamming filtering used to produce the 2-km posting product already attenuates the scales affected by the OBP ($<2\mathrm{ km}$) \citep{PeralOBP_21}. We ignore this effect for simplicity.

To simulate the 2-km posting SWOT product, we follow the standard SWOT post-processing method. The SWOT 2-km posting product is generated from the 250-m posting observations using a 17-point two-dimensional Hamming filter, which has a half-power cutoff wavelength of 6.28 km \citep{Chelton_24}. We mirror this process by convolving our previously generated 250-m posting Gaussian random fields with the two-dimensional Hamming filter and then sampling the data every 2~km. Solid lines in Fig. \ref{fig:Hamm_red} show the along-track spectra of the resulting 2-km posting data. The spectra now appear {red and power-law-like} at small scales. 
{Based on visual estimation}, the transition from the steep to the flat portion of the spectra varies with the noise level, {ranging from about 30~km for $\text{SWH}= 6\text{ m}$ to about 15 km for $\text{SWH}= 2\text{ m}$}. This shows that these transition scales depend on the ocean signal and noise level, which are all synthetic in our experiments. {In observations}, they will depend on geographic and temporal variability in real SWOT data.
Additionally, fields with smaller noise magnitude appear to have steeper spectral slopes, {indicating that the apparent small-scale slopes of the observed SWOT 2-km along-track spectra can} depend on the SWOT measurement noise levels.

It is worth exploring the reasons behind the {apparent red power-law spectra at small scales, which can be separated into three different effects. }

First, the noise floors in the 2-km posting spectra are lower than those in the 250-m posting spectra. This is well explained in prior literature (\citet[Appendix E]{CheltonEtAl_19}, \citet{WangEtAl_19}, \citet[Appendix A]{CheltonEtAl_22}, and \citet[Fig. 1]{Chelton_24}), and Fig. \ref{fig:Hamm_red} reproduces this known effect by comparing one-dimensional along-track filtering (dashed line) and two-dimensional filtering (solid line) for the same {synthetic SSH and noise corresponding to} $\text{SWH}=6\text{ m}$. 
{The combination of this lower noise floor and the steep ocean dynamical spectrum produces a redder spectrum (Fig.~\ref{fig:Hamm_red}, solid lines) than one-dimensional filtering does (Fig.~\ref{fig:Hamm_red}, dashed line). }

{
Second, the noise floor in the 2-km posting spectra is red over the smallest resolved scales because the 2-km grid resolves part of the Hamming filter’s spectral roll-off.
The 4-km Nyquist wavelength is smaller than the 6.28 km half-power filter cutoff wavelength of the 17-point Hamming filter. Hamming filtering already attenuates small scales starting at 15 km (see Fig.~1 in \citet{Chelton_24} and the black dashed line in Fig. \ref{fig:GM}). 
A closer match between the filter scale and the Nyquist wavelength could produce spectra that are less prone to visual misinterpretation.
Both of these behaviors were expected from \citet{Chelton_24}. 
}

Third, aliasing of the 2-km posting along-track spectra boosts the energy {near the Nyquist wavelength ($<10$ km).}
reshaping the spectrum so that it appears more like a red power law down to the Nyquist wavenumber
The dash-dotted line in Fig. \ref{fig:Hamm_red} is the along-track spectrum without aliasing. We achieve this by calculating the spectra {of the Hamming-filtered field at} 250-m posting but discarding the high-wavenumber portion. {The small-scale portions of these spectra then roll off in accordance with the Hamming filter's spectral response \cite[Fig. 1]{Chelton_24}.} This roll-off contrasts with the 2-km spectra, which continue the power-law-like behavior {up to the Nyquist wavenumber (solid lines). This is clear} between 4 and 8 km, where the aliased spectra {visually exhibit} slopes steeper than $-1$, even if the unfiltered signal contains only white noise (see the black dashed line in Fig. \ref{fig:GM}). This results from aliasing the Hamming filter's spectral roll-off.

{
\subsection{Analytical relation between the 250-m and 2-km-posting spectra}\label{sec:theory}
The along-track spectra can be calculated analytically, without Monte Carlo simulations, providing further insight into how filtering and sampling spatially white measurement noise can produce apparently red along-track spectra.\par

Let $x$ and $y$ denote the cross-track and along-track coordinates in units of kilometers, respectively, and let $k$ and $\ell$ denote the corresponding wavenumbers, in units of cycles per kilometer (cpkm). 
We denote by $P_2(k,\ell)$ the two-dimensional spectrum of the generated 250-m field (with units of $\text{ cm}^2\text{cpkm}^{-2}$), which combines the spectrum of the Mat\'ern ocean signal with that of white measurement noise. Let $h_{pq}$ be the normalized two-dimensional 17-point Hamming filter. With grid spacing $\Delta=0.25$~km, the filter's spatial Fourier transfer function is
\begin{align}
\hat{h}(k,\ell)=\sum_{p=-8}^{8}\sum_{q=-8}^{8}h_{pq}\exp\left[-2\pi i\Delta(pk+q\ell)\right].
\end{align}
The first step is two-dimensional convolution filtering, which multiplies the 250-m ocean-signal-plus-noise spectrum by the filter power response \citep[Appendix E]{CheltonEtAl_19}:
\begin{align}
P^\mathrm{filtered}_{2}(k,\ell)=|\hat{h}(k,\ell)|^2P_2(k,\ell).
\end{align}
The second step is to obtain the filtered along-track spectrum by integrating over all resolved cross-track wavenumbers \citep{Batchelor_53, BuhlerEtAl_14,SkinnerEtAl_25}:
\begin{align}
P^\mathrm{filtered}_{1}(\ell)=\int_{-2}^{2}P^\mathrm{filtered}_{2}(k,\ell)\,\mathrm{d}k, \\\qquad
-2\leq\ell<2\ \mathrm{cpkm}.\nonumber
\end{align}
This is the dealiased reference spectrum (cf. dash-dotted lines in Fig. \ref{fig:Hamm_red}). The final step is sampling every 2~km, which folds filtered spectral branches outside the 2-km posting Nyquist interval, \(-0.25\leq\ell<0.25\ \mathrm{cpkm}\), back into that interval. The resulting spectrum is \citep{BlackmanTukey_58b, Dutykh_16}
\begin{align}
P^{2\mathrm{km}}_{1}(\ell)
=
\sum_{\substack{m\in\mathbb{Z}\\-2\leq\ell+0.5m<2}}
P^\mathrm{filtered}_{1}(\ell+0.5m),\\
\qquad
-0.25\leq\ell<0.25\ \mathrm{cpkm}.\nonumber
\end{align}
The $m=0$ term is the directly resolved branch, while the $m\neq0$ terms are the high-wavenumber contributions folded into the 2-km Nyquist interval. This explains why the aliased spectrum has more energy than the dealiased spectrum near the Nyquist wavenumber. $P^{2\mathrm{km}}_{1}(\ell)$ is the theoretical two-sided along-track spectrum of the 2-km posting data. The solid lines in Fig. \ref{fig:Hamm_red} are the corresponding one-sided spectra obtained from the Monte Carlo simulations. Thus, at positive wavenumbers, the theoretical spectrum must be doubled, $2P^{2\mathrm{km}}_{1}(\ell)$, before comparison with those lines. The theoretical and Monte Carlo spectra are visually indistinguishable (comparison not shown).
}

\subsection{Accounting for cross-track variation in the noise level}\label{sec:cross}
SWOT's noise is known to vary in the cross-track direction \citep{PeralEtAl_24,KacheleinEtAl_25}, {adding complexity to understanding the impact of white noise on SWOT's smallest scales. We explore the effect of cross-track-dependent noise using the cross-track noise amplitude profile based on preliminary data reported in Fig. 30a of \citet{PeralEtAl_24}. Generally, for each 50-km SWOT swath, measurement noise is lowest near the center and higher at the points closest to and farthest from nadir.}

Here, we scale the noise variance across track while keeping the total noise variance unchanged. We then repeat the convolution of the data with the same two-dimensional Hamming filter as before (Section ~\ref{sec:experiments}\ref{sec:hamm_red}), but here we discard the along-track lines within 2.5 km of the {swath} boundaries to ensure that the Hamming filter does not use padded data.\par

\begin{figure*}
    \centering
    \includegraphics{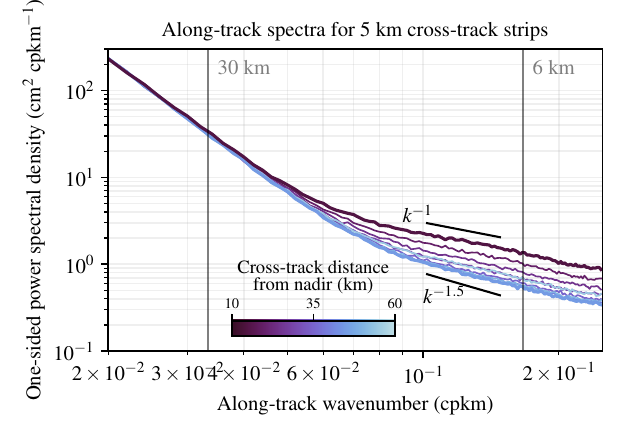}
    \caption{Along-track spectra of the 2-km posting product with cross-track-varying noise. {Spectra are averaged over 5~km cross-track slices and a $2\times 10^3$-member ensemble.} The prescribed noise follows the $\text{SWH}=4\text{ m}$ case in Fig. 30a of \citet{PeralEtAl_24}.
    There are nine lines in total since we do not use the data within 2.5 km of the swath edges, where the Hamming filter needs padded data. {The thicker line with the highest noise spectral level is the 5-km bin centered at 15 km cross-track, and the thicker line with the lowest noise spectral level is the 5-km bin centered at 45 km cross-track. 
    The black solid power-law lines are for visual reference over the range where they are plotted.}}
    \label{fig:Hamm_crosstrack}
\end{figure*}

{Figure \ref{fig:Hamm_crosstrack} shows the along-track spectra for the $\text{SWH}=4\text{ m}$ case, averaged within 5-km cross-track bins and over an ensemble of $2\times 10^3$ realizations to reduce sampling uncertainty.} 
{The track closest (15 km) to nadir has the highest noise magnitude
and hence has the shallowest slope at small scales. Tracks near the swath center (45 km from nadir) have the lowest noise level, and hence the steepest slopes \citep[Fig. 30a]{PeralEtAl_24}.
The cross-track-dependent noise level leads to cross-track-varying apparent small-scale slopes in the observed SWOT 2-km along-track spectra, consistent with the behavior in Fig.~\ref{fig:Hamm_red}. 
Based on visual comparison of spectra centered at 15 km and 45 km from nadir in Fig. \ref{fig:Hamm_crosstrack}, the slope can vary substantially across the swath, from $-1$ to $-1.5$, even for a uniform SWH, providing ample variation in spectral behavior. Note that the spectra averaged across the full swath are indistinguishable from those in the constant-noise case (Fig. \ref{fig:Hamm_red}), suggesting that studying only the cross-track-averaged spectra obscures the details of the cross-track behavior of SWOT noise.}

\subsection{{The impact of an internal gravity wave (IGW) continuum on filtered small-scale spectra}}
\label{sec:GM}
{We recognize that the resulting spectral slopes in Section \ref{sec:experiments}\ref{sec:cross} are flatter than those reported by \citet{KacheleinEtAl_25}, likely due to differences in the selected ocean signal. Our analysis in Section \ref{sec:experiments}\ref{sec:cross} assumes a steep balanced dynamical spectrum, whereas \citet{KacheleinEtAl_25} likely includes substantial IGW energy typical for the California region \citep{QiuEtAl_18}. This suggests that both the noise level and the assumed ocean signal matter in establishing red spectral slopes. Here, we explore the impact of a Garrett--Munk continuum \citep[GM,][]{GarrettMunk_72} as the ocean signal on noisy SWOT observations.}

SWOT might resolve parts of the GM spectrum for certain times and locations, though the magnitude of the associated SSH signature remains highly uncertain \cite[and references therein]{SamelsonFarrar_24}. We prescribe the two-dimensional GM spectra to {have the same form as} Eqs. (22) and (24) of \citet{SamelsonFarrar_24},
\begin{align}
    \frac{|K|^2}{(|K|^2+C)^2}\frac{1}{|K|},
\end{align}
where $C$ is a constant that determines the wavenumber of the spectral maximum, {$|K|=\sqrt{k^2+\ell^2}$} is the magnitude of the wavenumber vector, and the $1/|K|$ factor accounts for the conversion from the isotropic spectra to the two-dimensional spectra. The spectral shape is admittedly too simple, as it corresponds to the SSH spectrum for one vertical mode. It is nevertheless adequate for our purposes, since our primary interest is in the power-law behavior at high wavenumbers. To account for the uncertainty in our understanding of the SSH manifestation of the GM continuum, we consider spectral amplitudes scaled by factors of 1, 3, 10, and 30, roughly equidistant in log space. 
We here choose the noise level corresponding to an SWH of 3~m. This set of experiments implicitly explores other noise levels, since the results only depend on the magnitude of the ocean dynamical signal relative to the noise floor. {We estimate the spectra by averaging over all tracks across the generated swath for 200 independent ensemble realizations.}

The dotted lines in Fig. 3 show the along-track spectra of the synthetic GM-like signal and noise separately with no filtering (250-m posting data) \cite[Fig.~3]{SamelsonFarrar_24}. As a visual reference, the SWOT prelaunch noise requirement is also added in solid gray ({\citet{SWOT_sci_req}}, \citet[eq. (15)]{SamelsonFarrar_24}).
{At scales smaller than~100 km, the GM along-track spectra recover a $k^{-2}$ spectral slope, consistent with the noise-free GM prediction. }
{At scales larger than~100 km}, the along-track spectra are white, differing from the blue behavior in the isotropic GM spectrum shown in \citet{SamelsonFarrar_24}, Fig.~2a, {due to the superposition of multiple waves propagating in directions different from the along-track direction \citep{SamelsonFarrar_24}.} 
This difference between the isotropic spectra and along-track spectra is another example of how along-track spectra can significantly distort the two-dimensional picture. 

\begin{figure*}
    \centering
    \includegraphics{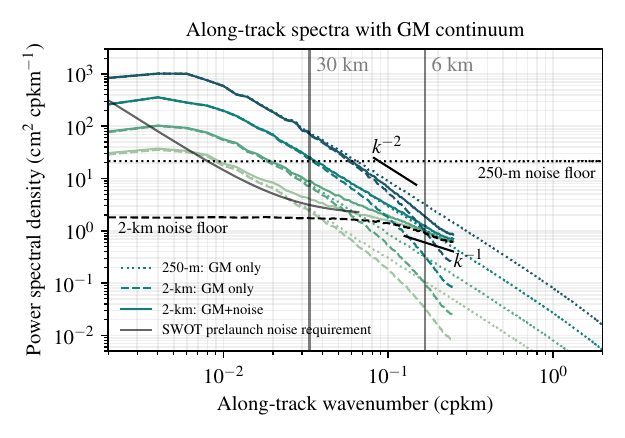}
    \caption{Along-track spectra for experiments using the Garrett--Munk (GM) spectra as the ocean signal. {Spectra are averaged over cross-track position and a 200-member ensemble.} We include white noise corresponding to an $\text{SWH of } 3\text{ m}$. 
    The colors represent different (artificial) GM levels, with the spectral magnitudes scaled by factors of $3$, $10$, and $30$ relative to the lowest level (light to dark green). 
    The gray solid line shows the SWOT prelaunch noise requirement as a visual reference.
    All dotted lines are along-track spectra of the 250-m posting data before Hamming filtering. The colored dotted lines show the GM-only part, and the black dotted line shows the noise-only part. 
    All solid and dashed lines are along-track spectra of the 2-km posting data after two-dimensional Hamming filtering. 
    The colored dashed lines show the GM-only spectra, the black dashed line shows the noise-only spectrum, and the colored solid lines show the total spectra (GM+noise). {The black solid power-law lines are for visual reference over the range where they are plotted.}}
    \label{fig:GM}
\end{figure*}

{Next, we look at the along-track spectra of the 2-km posting data of the GM ocean signal with noise (GM+noise, Fig. \ref{fig:GM} solid colored lines) and without noise (GM-only, Fig. \ref{fig:GM} dashed colored lines).}
{The GM+noise spectra visually exhibit} spectral slopes steeper or shallower than $-2$, depending on the relative strength of the GM continuum and the noise {(Fig.~\ref{fig:GM} solid colored lines)}. 
{They can deviate from $k^{-2}$ behavior at scales as large as 50 km, for the case with the lowest signal-to-noise ratio (Fig.~\ref{fig:GM}, the lightest-colored solid line).}

{In contrast, the GM-only} along-track spectra (dashed colored lines)
do not exhibit a clear power-law regime {after two-dimensional Hamming filtering and instead become substantially} steeper at high wavenumbers.
{This shows caution is necessary when} interpreting a $k^{-2}$ slope in SWOT 2-km along-track spectra as a direct signature of the GM continuum.

\section{Discussion and Conclusion}\label{sec:conc}
{We reproduced the transition to apparent power-law-like along-track spectra at small scales ($< 50$ km) with slopes visually} close to or steeper than $-1$ using three ingredients: a steep large-scale spectrum, two-dimensional spatially uncorrelated (white) noise {at 250-m posting}, and standard SWOT 2-km filtering and posting. 
Our analyses show that a lower noise level relative to the ocean signal produces steeper apparent slopes at small scales
(Figs.~\ref{fig:Hamm_red}, \ref{fig:Hamm_crosstrack}, and \ref{fig:GM}), {qualitatively} consistent with analyses of observed SWOT 2-km along-track spectra \citep{ZhangCallies_25a,KacheleinEtAl_25}. {We also find that in the synthetic cases examined here the visually determined scale of transition to these noise-affected spectra can range from 15 to 50 km, with the largest transition scale appearing for the case with high noise and a low-amplitude GM ocean spectrum.}

These noise-model experiments help to clarify the interpretation of SWOT data. 
{Apparent small-scale red slopes in the along-track spectra do not, by themselves, establish either red measurement noise or a distinct oceanic power-law regime.}
Rather, they can result from white noise, two-dimensional Hamming filtering, aliasing due to the 2-km posting, and computing one-dimensional along-track spectra from the filtered two-dimensional field, as it is common practice for spectral analysis of SWOT data.
Compared to studies that model SWOT noise as red in two dimensions, our interpretation is more consistent with current understanding of SWOT measurement noise and the SWOT 250-m posting data \citep{PeralEtAl_24,HuangWang_26}.
{While this is not the only mechanism that could produce red spectra at small scales}, we suggest the outlined noise model as a baseline null hypothesis, against which interpretations of ocean physics in SWOT data should be {tested} when analyzing { transition scales and spectral slopes of} one-dimensional spectra from two-dimensional filtered data.\par

In conclusion, our results call for caution when interpreting one-dimensional along-track spectra at {submesoscales $(O(10-100\text{ km}))$}. The apparent noise floor, {the transition scale to flatter spectra, and the spectral} slope depend strongly on the {detailed properties of the ocean physics signal and noise}, as well as the two-dimensional filter applied. Additionally, it might be beneficial to use two-dimensional diagnostics on the \textit{unsmoothed} 250-m product when targeting submesoscale signals. Careful modeling of the SWOT noise has direct implications for {processing and interpretation of SWOT data (e.g., \cite{XiaoEtAl_23,TreboutteEtAl_23,ZhangCallies_25a,SkinnerEtAl_25,BertrandEtAl_26,TatsuSWOT}), because} the SSH noise model sets key likelihood weights {in optimization, denoising, and interpolation methods}, and hence has a substantial impact on the output \citep{Wunsch_96,Stein_99a}. 

Finally, a global time-varying {quantification of the noise-affected spectra} still requires better constraints on both ocean signal variability and SWOT's (linear and nonlinear) noise amplitude, which is correlated with properties of ocean surface waves \citep{PeralEtAl_15,WangEtAl_19,Bohe_23,PeralEtAl_24,HuangWang_26, ArdhuinEtAl_26,OsinaEtAl_26}. The SWH observed from SWOT shows substantial submesoscale variability in the sea state, driven by wave-current interactions, surface winds, and wave groups \citep{VillasBoasEtAl_25a}. Sea state variability might manifest as variability in SWOT SSH noise {magnitude}, potentially at meso-- and submesoscales. In particular, the spectral convolution in the nonlinear noise model of \citet{PeralEtAl_15} is related to wave groups and the surface wave envelope spectrum \citep{DeCarloEtAl_23, ArdhuinEtAl_26}. 
Wave-current interactions suggest that SSH noise magnitude may correlate with the true SSH signal at submesoscales. Future work could explore using two-dimensional SWOT observations of SWH to inform estimates of SSH measurement noise, with particular focus on meso-- and submesoscale variability \citep{BoheEtAl_25,HoEtAl_25}. 

\clearpage
\acknowledgments
We thank Tatsu Monkman for sustained discussion about SWOT SSH noise. We also acknowledge Tom Farrar, Jinbo Wang, and Shafer Smith for providing feedback on draft versions of this paper. We do not imply their endorsement.

{We thank the Editor Gilberto Javier Fochesatto, reviewer Dudley Chelton, and two additional reviewers for providing reviews that improved the quality of the paper.}

RSD, LL, MH, and ABVB are supported by NASA award 80NSSC24K1647 through the SWOT Science Team. ABVB received additional support from the ONR MURI program (Grant N00014-24-1-2554). LL received support from NASA award 80NSSC23K0985. AI tools were used to develop the software used in this paper. The authors are fully responsible for the scientific content and integrity of this paper.

\datastatement
The code for the experiments of this paper is available at \url{https://github.com/Empyreal092/SWOT_noise_1D2D}. A final version of the code will be released with a DOI upon acceptance.

{
\appendix[A]
\appendixtitle{Scaling the white-noise spectrum to 250-m posting}

For completeness, here we derive the factor $2/(1/15)$ used in Eq.~\eqref{eq:noise_floor} to
scale the SWOT prelaunch white-noise requirement specified for wavenumbers less than $1/15$ cpkm to the 250-m posting. These derivations appeared in earlier literature, including \citet[Section 2e]{WangEtAl_19} and \citet[Appendix E]{CheltonEtAl_19}. 

Assume spatially white, two-dimensional noise with spectral density
\begin{equation}
P_2(k,\ell)=N_0,
\end{equation}
where $k$ and $\ell$ are the cross-track and along-track wavenumbers,
respectively and $P_2(k,\ell)$ the two-dimensional spectrum (with units of $\text{ cm}^2\text{cpkm}^{-2}$). For a separable filter, $\hat{h}(k,\ell)=\hat{h}_x(k)\hat{h}_y(\ell)$, the
filtered along-track spectrum is the cross-track integral of the filtered
two-dimensional spectrum:
\begin{equation}
\begin{aligned}
P_{1}^\mathrm{filtered}(\ell)
&= \int
|\hat{h}_x(k)|^2 |\hat{h}_y(\ell)|^2 P_2(k,\ell)\,\mathrm{d} k \\
&= N_0 |\hat{h}_y(\ell)|^2
\int|\hat{h}_x(k)|^2\,\mathrm{d} k.
\end{aligned}
\label{eq:separable-white-noise}
\end{equation}
Thus, cross-track filtering sets the magnitude of the along-track white-noise
spectrum.

Now take the cross-track filter to be an ideal rectangular low-pass filter in
wavenumber space,
\begin{equation}
\hat{h}_x(k)=
\begin{cases}
1, & |k|\leq k_c,\\
0, & |k|>k_c.
\end{cases}
\end{equation}
Equation~\eqref{eq:separable-white-noise} then becomes
\begin{equation}
P_{1}^\mathrm{filtered}(\ell)=2k_c N_0 |\hat{h}_y(\ell)|^2.
\label{eq:ideal-low-pass-white-noise}
\end{equation}
At along-track wavenumbers where $|\hat h_y(\ell)|=1$, the one-dimensional noise level is proportional to the retained cross-track bandwidth,
$2k_c$. Consequently, scaling from the requirement bandwidth
$k_c=1/15$ cpkm to the 250-m posting Nyquist wavenumber $k_c=2$ cpkm
gives
\begin{equation}
\frac{P_{1}^{250\,\mathrm{m}}}{P_{1}^\mathrm{req}}
=
\frac{2(2)}{2(1/15)}
=
\frac{2}{1/15}.
\end{equation}
Thus, this factor accounts for the larger cross-track wavenumber range
of white noise in 250-m posting.
The one-sided power spectral densities plotted in the paper are twice the corresponding two-sided power spectral densities at positive wavenumbers.
}

\bibliographystyle{ametsocV6}
\bibliography{citation}

\end{document}